\documentclass[12pt]{article}
\usepackage{graphics,color}
\begin{document}
\thispagestyle{empty}
\noindent\
\\
\\
\\
\begin{center}
\large \bf  Composite Leptons, Quarks and Weak Bosons
\end{center}
\hfill
 \vspace*{1cm}
\noindent
\begin{center}
{\bf Harald Fritzsch}\\
Department f\"ur Physik\\ 
Ludwig-Maximilians-Universit\"at\\
M\"unchen, Germany \\
\end{center}

\begin{abstract}

The leptons, quarks and weak bosons are bound states of new constituents. An isosinglet weak boson should exist, the $X$ - particle, which might have a  mass of about 400 GeV and a width of about 20 GeV. 

\end{abstract}

\newpage

In the Standard Theory of the electroweak interactios, based on the gauge group SU(2) x U(1), the universality of the weak interactions follows from the underlying gauge symmety. The masses of the weak bosons, the leptons and the quarks are generated by the spontaneous symmetry breaking. They depend on unknown coupling constants and cannot be calculated.\\ 

The masses of the bound states in confining gauge theories are generated by another mechanism, which is due to the confinement property  of the non-Abelean gauge force. In the theory of Quantum Chromodynamics the masses of the hadrons can be calculated. They are given  by the field energy of the confined gluons and quarks and are proportional to the QCD mass scale $\Lambda_c$, which has  to be measurend in the experiments: $\Lambda_c = 332 \pm 17 $  MeV.\\

In the Standard Theory the physics of the six leptons and the six quarks depends on 20 free prameteres: twelve masses of the leptons and quarks and eight parameters of the flavor mixing - six mixing angles and two phase parameters, which describe the CP violation. Perhaps these twenty parameters can be calculated in the future, when details about the physics beyond the Standard Theory are known. \\

Perhaps the twenty parameters indicate, that the leptons, quarks and weak bosons are not pointlike, but composite systems. Especially we shall assume, that inside the leptons and quarks are two pointlike particles, a fermion and  a scalar. They are called "haplons" ("haplos" means "simple" in the Greek language). The haplons are confined by massless gauge bosons, which are described by a gauge theory similar to quantum chromodynamics.  We call these gauge bosons "hypergluons" , and the associated quantum numbers "hypercolors". The gauge theory is called "Quantum Haplodynamics" (QHD).\\

We assume that the masses of the weak bosons are also generated dynamically. This is possible, if the weak bosons are not elementary gauge bosons, but bound states of haplons, analogous to the $\rho$-mesons in QCD. In 2012 a scalar "Higgs" boson with a mass of 125 GeV has been discovered at the LHC (ref.(1,2)). This new boson is an excitation of the $Z$-boson.\\

The weak bosons consist of two haplons, a fermion and its antiparticle (see also ref.(3,4,5,6,7)). The $QHD$ mass scale is given by a mass parameter $\Lambda_h$, which determines the size of the weak bosons.\\ 

The haplons are massless and interact with each other through the exchange of massless hypergluoms. The number of hypergluons depends on the gauge group, which is unknown. We assume, that it is the same as the gauge group of $QCD$: $SU(3)$. Thus the binding of the haplons is due to the exchange of eight massless hypergluons. \\ 

Two types of fermions are needed as constituents of the weak bosons, denoted by $\alpha$ and $\beta$. The electric charges of these two haplons are (+1/2) and (-1/2). The electric charges of the scalar haplons are (-1) for the haplon inside the leptons and (+1/6) for the three scalar haplons inside the quarks. The sum of the electric charges of the four scalar haplons is zero.\\

The three weak bosons have the following internal structure:
\begin{eqnarray}
W^+ = (\overline{\beta} \alpha),~~ W^- = (\overline{\alpha} \beta),~~W^3 =\frac{1}{\sqrt{2}} \left( \overline{\alpha} \alpha - \overline{\beta} \beta \right).
\end{eqnarray}

The $QHD$ mass scale can be estimated, using the observed value of the decay constant of the weak boson. This constant is analogous to the decay constant of the $\rho$-meson, which is directly related to the $QCD$ mass scale. The decay constant of the weak boson is given by the mass of the weak boson and the observed value of the weak mixing angle:
\begin{eqnarray}
F_{\rm W}=\sin\theta_{\rm W}\cdot \frac{M_{\rm W}}{e}.
\end{eqnarray}

We find $F_{\rm W} \simeq 0.125 ~{\rm TeV}$. Thus the $QHD$ mass scale $\Lambda_h$ is probably about 0.2 TeV, thousand times larger than the $QCD$ mass scale (see also ref. (7,8,9)). Due to the uncertainties of these estimates we expect that $\Lambda_h$ is in the range beteen 0.2 TeV and 1 TeV.\\

The haplon wave function of the neutral weak boson is given by: 
\begin{eqnarray}
W^3 =\frac{1}{\sqrt{2}} \left( \overline{\alpha} \alpha - \overline{\beta} \beta \right).
\end{eqnarray}

In our theory there must also exist an isoscalar particle with the wave function:
\begin{eqnarray}
X=\frac{1}{\sqrt{2}} \left( \overline{\alpha} \alpha + \overline{\beta} \beta \right).
\end{eqnarray}

According to the results of the LHC-experiments this particle $X$ cannot have a mass  similar to the mass of the neutral weak boson. The mass must be much larger than the mass of the $Z$ - boson. This is possible - the current, associated with the 
$X$ - particle, has an anomaly, generated by the interaction of the haplons with the hypergluons. Due to this anomaly the mass of the $X$ - particle is much larger than the mass of the  $Z$ - boson. As an example we shall assume, that the mass of the $X$ - particle is about 400 GeV. This mass could also be much larger, e.g. close to 800 GeV.\\

We estimate the cross section for producing the $X$ - particle at the LHC to about $3~nb$. The partial width of the decay of the $X$ - particle
into a pair of muons is estimated to about 0,4 GeV. The $X$ - particle can also decay into a pair of electrons, a pair of neutrinos, 
a quark and its antiquark, a pair of weak bosons and a pair of weak bosons and a photon. The total width of the $X$ - particle 
is estimated to about 20 GeV. The best way to observe the $X$ - particle is to observe the decay into a pair of muons, which would allow to measure the mass of the $X$ - particle. \\

The $X$ - particle is produced by the direct interaction of a quark and an antiquark. It cannot be produced by the interaction of two gluons. Thus it would be useful to  change the LHC and to arrange interactions of protons and antiprotons. A similar change was made years ago with the SPS accelerator at CERN. This change was important for the discovery of the weak bosons.  \\

There is a good chance that the  $X$ - particle is observed soon in the LHC-experiments. If this is the case, our hypothesis 
about the internal structure of the leptons, quarks and the weak bosons would be correct.


\begin{thebibliography}{99}
 
\bibitem{1} G. Aad et al. (Atlas collaboration) Phys.Lett. B 716, 1 (2012).

\bibitem{2} S. Chatrchyan et al. (CMS collaboration) Phys. Lett. B 716, 30 (2012).

\bibitem{3}  H. Fritzsch and G. Mandelbaum, Phys. Lett. B102 (1981) 319; \\
    Phys. Lett. B 109 (1982) 224.

\bibitem{4}  R. Barbieri, R. Mohapatra and A. Masiero, Phys. Lett. B 105
(1981) 369.

\bibitem{5}  H. Fritzsch. D. Schildknecht and R. Kogerler, Phys. Lett. B 114 (1982)
157.

\bibitem{6}  L. F. Abbott and E. Farhi, Phys. Lett. B 101, 69
(1981).

\bibitem{7} H. Fritzsch, Modern Physics Letters A, Vol. 31,  No. 20 (2016).

\bibitem{8} H. Fritzsch, Mod. Phys. Lett. A26, 2305 (2011).

\bibitem{9} H. Fritzsch, Mod. Phys. Lett. A, Vol. 31, No. 20 (2016), 1630019.


\end{thebibliography}
\end{document}